# Specific heat in $KFe_2As_2$ in zero and applied magnetic field

J. S. Kim[1], E. G. Kim[1], G. R. Stewart[1], X. H. Chen[2] and X. F. Wang[2]

[1]Department of Physics, University of Florida, Gainesville, FL 32611
[2]Hefei National Laboratory for Physical Science at Microscale and Department of Physics, University of Science and Technology of China, Hefei, Anhui 230026, China

**Abstract:** The specific heat down to 0.08 K of the iron pnictide superconductor $KFe_2As_2$ was measured on a single crystal sample with a residual resistivity ratio of ~650, with a $T_c^{onset}$ determined by specific heat of 3.7 K. The zero field normal state specific heat divided by temperature, C/T, was extrapolated from above $T_c$ to T=0 by insisting on agreement between the extrapolated normal state entropy at $T_c$, $S_n^{extrap}(T_c)$, and the measured superconducting state entropy at $T_c$, $S_s^{meas}(T_c)$, since for a second order phase transition the two entropies must be equal. This extrapolation would indicate that this rather clean sample of $KFe_2As_2$ exhibits non-Fermi liquid behavior, i. e. C/T *increases* at low temperatures, in agreement with the reported non-Fermi liquid behavior in the resistivity. However, specific heat as a function of magnetic field shows that the shoulder feature around 0.7 K, which is commonly seen in $KFe_2As_2$ samples, is <u>not</u> evidence for a second superconducting gap as has been previously proposed but instead is due to an unknown magnetic impurity phase, which can affect the entropy balance and the extrapolation of the normal state specific heat. This peak (somewhat larger in magnitude) with similar field dependence is also found in a less pure sample of $KFe_2As_2$, with a residual resistivity ratio of only 90 and $T_c^{onset}$=3.1 K. These data, combined with the measured normal state specific heat in field to suppress superconductivity, allow the conclusion that an increase in the normal state specific heat as T→0 is in fact not seen in $KFe_2As_2$, i. e. Fermi liquid behavior is observed.

**I. Introduction**

As an endpoint in the $Ba_{1-x}K_xFe_2As_2$ system, in which high temperature iron compound superconductivity was first discovered[1] at x=0.4 in the 122 structure, $KFe_2As_2$ has been the subject of rather intense interest. The problem of making good samples that represent 'intrinsic' behavior in this system has been a continuing struggle. However, it seems clear that the superconductivity in $KFe_2As_2$ is unusual in several aspects. First, $KFe_2As_2$ shows clear evidence of nodal behavior based on the linear temperature dependence of the penetration depth[2], $\Delta\lambda(T)\sim T$, and on the field dependence of the thermal conductivity[3], $\kappa(H)\sim H^{1/2}$. Second, $KFe_2As_2$ exhibits non-Fermi liquid (nFl) behavior in its normal state electrical resistivity[3], $\rho=\rho_0 + AT^{1.5}$. Third, the specific heat discontinuity at $T_c$, $\Delta C$, does not follow the correlation first noted by Bud'ko, Ni and Canfield[4] ('BNC') and expanded upon by J. S. Kim et al.[5] that $\Delta C/T_c \sim 0.083 T_c^{1.9}$ for the FePn/Ch superconductors. As will be shown for the rather high quality sample in the present work, $\Delta C/T_c \sim 41$ mJ/molK$^2$ for $KFe_2As_2$ at $T_c^{midpoint}$=3.1 K while the BNC phenomenological correlation would predict $\Delta C/T_c \sim 0.7$ mJ/molK$^2$.

In the present work we argue that the straightforward extrapolation of the zero field normal state specific heat to match the measured superconducting state entropy gives incorrect evidence for non-Fermi liquid behavior in $KFe_2As_2$.  In an attempt to understand the lower temperature anomaly (~ 0.7 K) in the specific heat analyzed by others[6] in their samples of $KFe_2As_2$ in terms of a two band gap model, we measured the specific heat in magnetic field and found that the low temperature anomaly appears to be due to a magnetic impurity rather than to a smaller superconducting band gap.  This is further supported by specific heat data on a less pure comparison sample.  The field data also obviate the need for extrapolating the normal state data from above $T_c$ and, at least above the magnetic anomaly, show no increase in $C/T_{normal}$ as temperature is lowered.

## II.  Experimental

The high quality $KFe_2As_2$ single crystals were grown using KAs self flux.  The KAs flux was prepared by reacting stoichiometric quantities of K (in small pieces) with As powder at 200 °C for four hours.  Stoichiometric amounts of the elements together with the pre-reacted flux were then mixed as $KFe_2As_2$:KAs in a ratio of 1:6, placed in an alumina crucible and sealed in a stainless steel container.  The samples were heated to 930 °C and slow cooled to 500 °C.  The KAs flux was dissolved by washing in alcohol.  Resistive superconducting onset temperatures for three samples of $KFe_2As_2$ prepared in this manner varied between 4.2 and 4.4 K, with transition widths of 0.4-0.5 K.  Residual resistivity ratios ('RRR'), defined as

$\rho$(300K) divided by the resistivity extrapolated to T=0 from above $T_c^{onset}$, varied between 600 and 680.

$KFe_2As_2$ crystals of lower (~90) RRR were prepared using pre-reacted FeAs self flux. The FeAs flux was prepared by mixing together stoichiometric amounts of Fe and As powders, pressing into pellets, and reacting at 900 $^o$C for 10 hours. Stoichiometric amounts of the elements together with the reground FeAs flux were then mixed as $KFe_2As_2$:FeAs in a ratio of 1:2, placed in an alumina crucible and sealed in a Nb container. The samples were heated to 1150 $^o$C, held for five hours, and cooled at 5 $^o$C/hr to 500 $^o$C, and then cooled at 75 $^o$C/hr to room temperature. Samples were removed mechanically from the FeAs flux. Resistive superconducting onset temperatures for samples prepared using this method were typically 3.8 to 3.9 K, with transition widths of order 1.0 K. RRR values varied between 86 and 96, similar to values reported in the literature[3] for this preparation method.

Several small single crystals of total mass 6.29 mg of the high RRR sample/two single crystals of total mass 8.00 mg of the lower RRR sample were attached using GE 7031 varnish to our specific heat sample platform and measured in zero and applied field using standard techniques[7].

### III. Results and Discussion

The specific heat of the collage of high quality (RRR~ 650) single crystals in the form of flat platelets of $KFe_2As_2$ down to 0.08 K is shown in Fig. 1. A smooth extrapolation of the normal state data down to T=0 that gives the same normal state entropy, $S_n(T_c)$, as is measured in the superconducting state, $S_s(T_c)=\int C_s/T\, dT$ (integral is from 0 to $T_c^{onset}$), is shown that satisfies the requirement that $S_n(T_c)=S_s(T_c)$ for a second order phase transition. This extrapolation implies that $C_n/T$ *increases* as temperature is lowered past about 2.3 K. Such an increase in C/T at low temperatures, although only about 10 %, would normally imply a material is a non-Fermi liquid since for a Fermi liquid the electronic contribution to C/T, $\gamma$, is constant at low temperatures with $C/T=\gamma+\beta T^2$ and $\beta T^2$ is the lattice phonon contribution. However an exception to this is if there is a transition/anomaly at low temperatures that concentrates entropy over a narrow temperature range.

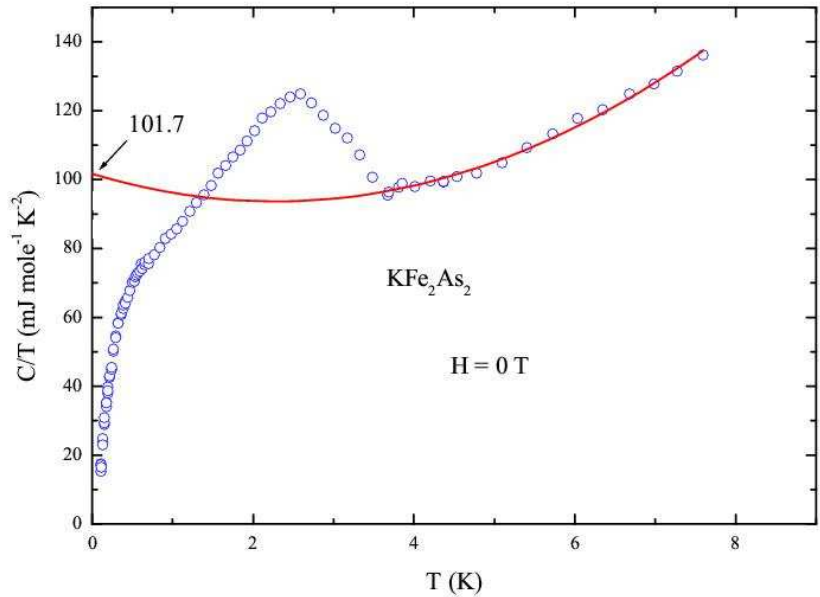

Fig. 1 (color online): Specific heat data in zero field of $KFe_2As_2$, with RRR ~ 650 down to 0.08 K. Extrapolation of the normal state data (red line) is drawn to match the normal and superconducting state entropies at $T_c$. As can been seen, this fit implies nFl behavior (by definition, in a Fermi liquid $\gamma$ is independent of temperature at low temperature) like observed in the heavy Fermion superconductor $UBe_{13}$, where $\gamma$ increases 20% below $T_c$ = 0.9 K. Note however the shoulder in the lower temperature specific heat data as discussed in the text. The rather large, ~ 1 K, transition width in this bulk specific heat measurement is unusual in such a large RRR material, and may imply a sensitivity of the superconductivity to rather low levels of defects.

The field dependence of the shoulder in the low temperature specific heat at around 0.7 K, which appears like the signature of a second, smaller superconducting gap as has been observed[8] in, e. g., $MgB_2$, is shown in Fig. 2. The field was applied in the ab-plane of the crystal collage, based on a report[9] that $H_{c2}(T=0)$~4.5 T for this field direction in a sample of $KFe_2As_2$, RRR=87, while $H_{c2}(T=0)$~1.3 T for H||c. As

the applied magnetic field increases, the suppression of the superconducting specific heat anomaly, $T_c^{onset}$=3.68 K in zero field, progresses such that in 4.5 T this anomaly can no longer be distinguished. As an aside, note that $H_{c2}(T=0)$ for the present sample must be larger than 4.5 T since C/T as T→0 is still falling sharply in this field

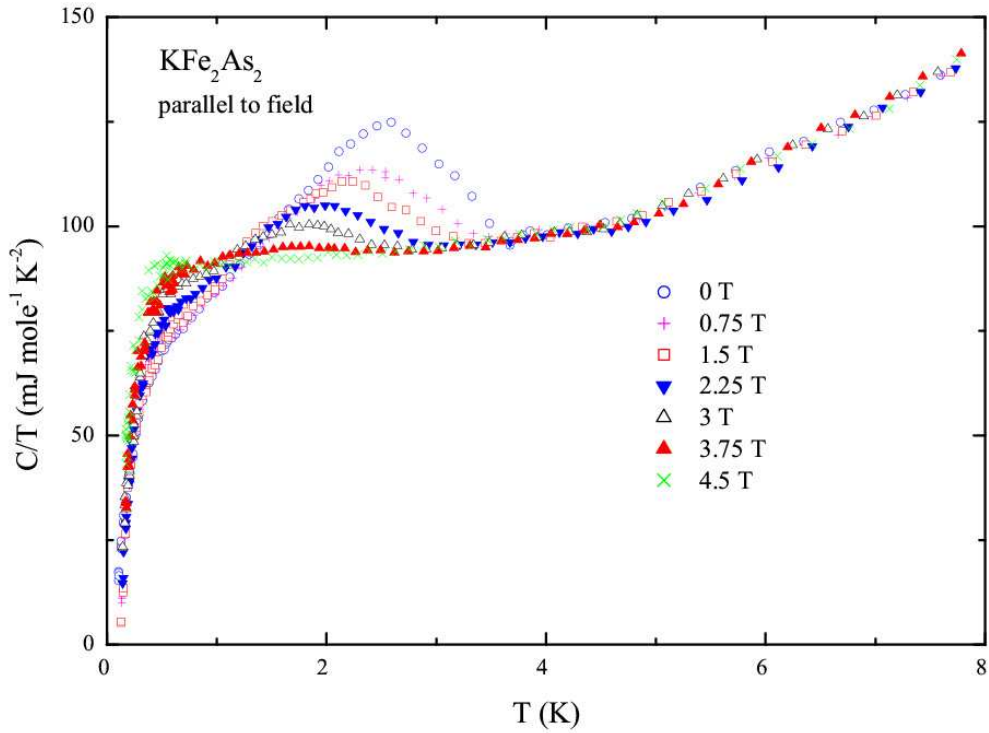

Fig. 2 (color online): Specific heat as a function of field for high quality single crystalline KFe$_2$As$_2$, with the field in the plane of the flat platelet crystals, i. e. in the ab plane. Note that although the superconducting transition is suppressed in temperature with increasing field, and is absent by $H_{c2}$ ~ 4.5 T, the 0.7 K shoulder becomes more pronounced with increasing field. The amount of entropy associated with this anomaly, presumably proportional to the concentration of the impurity phase that produces it, is only about 0.03% of Rln2. The specific heat discontinuity ΔC at the superconducting transition divided by $T_c$, $\Delta C/T_c$, is 41 mJ/molK$^2$ for an idealized transition at $T_c^{mid}$=3.1 K.

in Fig. 2. Plotting applied field vs $T_c^{onset}$ from the specific heat data from Fig. 2 in

Fig. 3 shows that $H_{c2}(T=0)$ for this clean sample of $KFe_2As_2$ is approximately 7.1 T for field in the ab-plane.

Focusing now on the anomaly at 0.7 K, its apparent non-superconducting nature (since it remains at approximately the same temperature as field in increased as shown in Fig. 2) provides an alternative explanation to the need for the extrapolated $C_{normal}/T$ in Fig. 1 to increase as T→0 to match the superconducting state entropy. Thus, the normal state C/T (as shown by the data in field in Fig. 2) just decreases monotonically with decreasing temperature (Fermi liquid behavior) and accretes the extra entropy from the impurity normal phase component (the 0.7 K anomaly) in order to match the superconducting state entropy (which of course also includes the anomaly.)  Following the 4.5 T data down to 0.8 K in Fig. 2 (i. e. just above the anomaly), it is apparent that the entropy-driven extrapolation in Fig. 1 is incorrect and in fact the normal state C/T continues to fall as temperature is lowered.

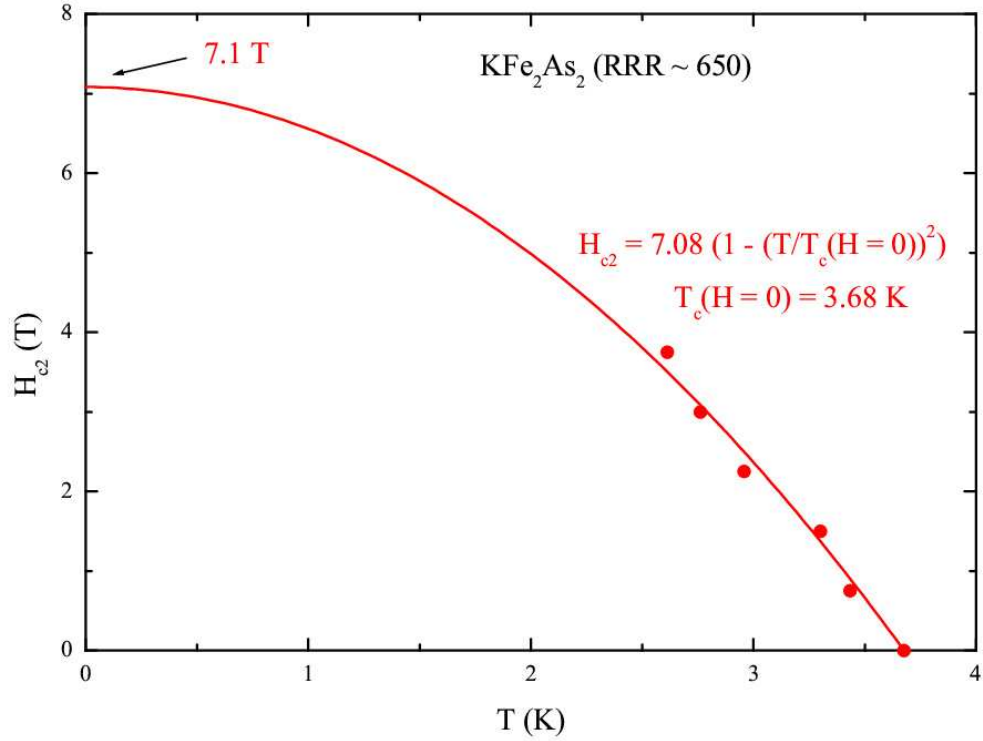

Fig. 3 (color online): Upper critical field as a function of temperature in high quality KFe$_2$As$_2$, field in the ab-plane. The red line extrapolation follows the function form H$_{c2}$ = H$_{c2}$(T=0)*(1-(T/T$_c$)$^2$), and the T$_c$'s used are T$_c^{onset}$ as determined from the bulk specific heat data in Fig. 2. Even if the T$_c$'s are determined from the position of the *peak* in the specific heat (which would correspond to the resistive transition being fully finished), an extrapolation of these T$_c^{peak}$ data (not shown) gives H$_{c2}$(T=0)=6.1 T. This is another example of sample dependent properties in KFe$_2$As$_2$, contrasting with the report of Terashima et al.[9] that H$_{c2}$(T=0)=4.5 T, H||ab, using resistive measurements of their RRR=87 sample of KFe$_2$As$_2$.

In order to further investigate this anomaly, Fig. 4 shows the specific heat in zero and applied field down to 0.4 K of a lower quality, RRR=90, sample. The low temperature anomaly is more prominent, containing six times the entropy of the corresponding anomaly in the high RRR sample, and more rounded, and occurs at the slightly higher temperature of ~ 1 K. Even up to 12 T the anomaly is only suppressed gradually to lower temperature, implying that the internal exchange fields are rather large. There are numerous examples where the exchange fields are much larger in energy than the ordering temperature of the magnetic transition, e. g. in CeRhIn$_5$ the temperature of the antiferromagnetic transition is 3.8 K and 45 T suppresses this only to 3.2 K[10]. Clearly, this anomaly is an important source of the entropy in the extrapolation of the normal state specific heat data below T$_c$.

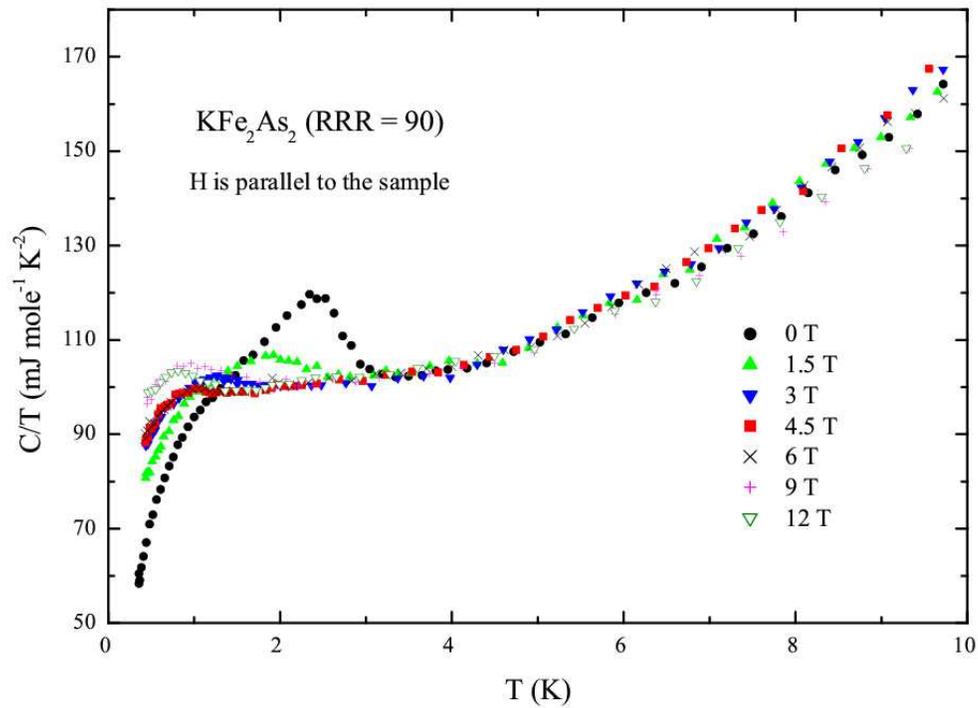

Fig. 4 (color online): Specific heat as a function of magnetic field, H ∥ ab plane, down to 0.4 K on a lower quality (RRR=90) single crystalline sample of KFe$_2$As$_2$ than shown in Figs. 1 and 2. The 4.5 and 6 T data are essentially identical below 1.5 K. An extrapolation (not shown) of the zero field normal state data to T=0 to match the superconducting state entropy gives γ=91 mJ/molK$^2$, with C$_{normal}$/T in this extrapolation decreasing monotonically as T→0.

## IV. Summary and Conclusions:

Specific heat in magnetic fields up to 12 T and to temperatures down to 0.08 K on two single crystal samples of $KFe_2As_2$ of differing qualities (RRR~650 and 90) shows that the anomaly around 0.7 K in the specific heat is due to a magnetic phase. Since the anomaly is enhanced in the lesser quality samply, we conclude that it is extrinsic as has been seen[11] in the specific heats of other iron pnictide superconductors at low temperatures. The specific heat measured in fields to suppress superconductivity behaves like a Fermi liquid, i. e. γ decreases monotonically with decreasing temperature, down to the magnetic anomaly. Sample dependence continues to play a role in the measured properties of this interesting compound, with the extrapolated $H_{c2}(0)$ for the RRR~650 sample for field in the ab plane being almost 60% larger than reported in the literature for an RRR=87 sample.

**Acknowledgements:** The authors gratefully acknowledge discussions with Eric Bauer, Peter Hirschfeld, Kevin Ingersent and Vivek Mishra. Work at Florida performed under the auspices of the United States Department of Energy, contract no. DE-FG02-86ER45268. Work below 0.4 K in magnetic field performed at NHMFL, Tallahassee which is supported by the United States National Science Foundation.